\documentclass[12pt]{article}

\usepackage{amssymb}
\usepackage{amsmath}

\makeatletter

\def\boldit#1{\mbox{\boldmath$#1$}}
\def\eqalign#1{\null\vcenter{\def\\{\cr}\openup\jot\m@th
 \ialign{\strut$\displaystyle{##}$\hfil&$\displaystyle{{}##}$\hfil
     \crcr#1\crcr}}\,}

\makeatother

\title{Canonical description of ideal magnetohydrodynamics and integrals of motion.}

\author{A.V. KATS \\
Usikov Institute for Radiophysics and Electronics National Academy of Sciences \\ of
Ukraine,
  61085, 12 Ak. Proskury St., Kharkiv, Ukraine, \\
e-mail:  avkats@akfirst.kharkiv.com; avkats@ire.kharkov.ua}

\begin{document}



\maketitle

\begin{abstract}

In the framework of the variational principle there are introduced canonical variables
describing magnetohydrodynamic (MHD) flows of general type without any restrictions for
invariants of the motion. It is shown that the velocity representation of the  Clebsch
type introduced by means of the variational principle with constraints is equivalent to
the representation following from the generalization of the Weber transformation for the
case of arbitrary MHD flows. The integrals of motion and local invariants for MHD are
under examination. It is proved that there exists generalization of the Ertel invariant.
It is expressed in terms of generalized vorticity field (discussed earlier by Vladimirov
and Moffatt (V.~A.~Vladimirov, H.~K.~Moffatt, J. Fl. Mech., {\bf 283}, pp.~125--139,
1995) for the incompressible case). The generalized vorticity presents the frozen-in
field for the barotropic and isentropic flows and therefore for these flows there exists
generalized helicity invariant. This result generalizes one obtained by Vladimirov  and
Moffatt in the cited work for the incompressible fluid. It is shown that to each
invariant of the conventional hydrodynamics corresponds MHD invariant and therefore our
approach allows correct limit transition to the conventional hydrodynamic case. The
additional advantage of the approach proposed enables one to deal with discontinuous
flows, including all types of possible breaks.

\end{abstract}

\section{Introduction.}

 \hskip\parindent

It is well-known that description of the solid media flows in terms of the canonical
(hamiltonian) variables is very useful and effective, see for instance \cite{Zh_Kuz_97,
GP93}. In terms of the hamiltonian variables it is possible to  deal with  all nonlinear
processes in unified terms  not depending on the specific problem related to the media
under investigation. For instance, all variants of the perturbation theory are expressed
in terms of different order nonlinear vertices, which along with the linear dispersion
relation contain the specific information relating to the concrete system under
investigation, cf. Refs. \cite{ZLF_92, KUZ_01}. In the problems of the nonlinear
stability investigations the conventional Hamiltonian approach based upon the
corresponding variational principle allows one to use the Hamiltonian along with other
integrals of motion (momentum, number of quasi-particles, topological invariants) in
order to construct the relevant Lyapunov functional, cf. Refs. \cite{Arn65, Abarb_83,
Lewis_86, Vl95, Vl96}. Therefore, it makes important the problem of introducing the
canonical variables and corresponding variational principle for the general type MHD
flows (i.~e., non-barotropic and including all types of the breaks possible for MHD) and
obtaining the complete set of the local invariants, see definition and discussions in
original papers \cite{MCTYan_82, STY_90, TYan_93, VTY_95} and in the recent review
\cite{Zh_Kuz_97}. As for the first item, the example of the variational principle
describing all possible breaks is presented in the recent work \cite{KATS_02}.

Here in the framework of some modification of the variational principle of the cited work
we examine the problem of the MHD invariants. Note  that the set of invariants for MHD
discussed in the literature till now is incomplete. It becomes evident if one takes into
account that for the vanishing magnetic field this set has to go over to the set of the
conventional hydrodynamic invariants. But this limit transition does not reproduce Ertel,
vorticity and helicity invariants existing for the hydrodynamic flows. For the particular
case of incompressible MHD flows generalized vorticity and helicity invariants were
obtained in the paper \cite{Vl95}. Below we show that the generalized vorticity and
helicity invariants exist also for compressible barotropic flows, and derive MHD
generalization for the Ertel invariant.

The plan of the paper is as follows. In  section \ref{VAR_PR} we briefly discuss
appropriate variational principle, introducing the Clebsch type velocity representation
by means of constraints and defining the canonical variables. In the following section
\ref{WEB_TR} we develop generalization of the Weber transformation and show that it leads
to the velocity representation, which is equivalent to the one following from the
variational principle under discussion. In section \ref{INT_MOTION} we examine MHD
integrals of motion, introducing `missing' MHD invariants, and discuss their
transformation properties relating to change of gauge. In section \ref{Conlusions} we
make some conclusions and formulate problems to be solved later.

\section{Variational principle and canonical variables.}\label{VAR_PR}

 \hskip\parindent

Let us briefly describe the variational principle and subsidiary variables describing
dissipation-free MHD. Starting with the standard Lagrangian density
\begin{equation}\label{Lagr_1A_Er}
  {\mathcal L} = \rho\frac{v^2}{2} -
\rho \varepsilon(\rho, s) - \frac{\mathbf{H}^{2}}{8\pi} \, ,
\end{equation}
where $\rho$, $s$ and $\varepsilon(\rho, s)$ present the fluid density, entropy and
internal energy, respectively, $\mathbf{H}$ denotes magnetic field, we have to include to
the action $\mathcal{A}$ the constraint terms. Then the action can be presented in the
form
\begin{equation}\label{Action_vol_Er}
\mathcal{A} =  \int d t L' \, , \qquad L' = \int d \mathbf{r} \mathcal{L}'  \, , \qquad
\mathcal{L}' = \mathcal{L} + \mathcal{L}_{c} \, ,
\end{equation}
where $\mathcal{L}_{c}$ is the part of the Lagrangian density respective for constraints,
\begin{equation}\label{Constr_8_Er}
 \mathcal{L}_{c} = \rho D\varphi + \boldit{\lambda} D\boldit{\mu} + \sigma D s  +  {\mathbf{M}} \cdot
  \left(
\frac{\partial \mathbf{A}}{\partial t} -  [{\mathbf{v}}, curl {\mathbf{A}} ]  + \nabla
\Lambda \right) + \frac{\mathbf{H} curl \mathbf{A}}{4 \pi} \, .
\end{equation}
Here $D = \partial_{t} + (\mathbf{v}\nabla)$ is substantial (material) derivative and
$\mathbf{A}$ is the vector potential.\footnote{ This form of the action slightly differs
from the one proposed in Ref. \cite{KATS_02}. The main difference consists in introducing
the vector potential for the magnetic field. Therefore, here the canonical pair is
$\mathbf{A}, \mathbf{M}$ instead of $\mathbf{H}, \mathbf{S}$, where $\mathbf{S} = curl
\mathbf{M}$. We do not deal here with the discontinuous flows and thus we omit the
surface term in the action. But adding corresponding surface term we can easily take the
breaks into account.} Including the last two terms into $\mathcal{L}_{c}$ allows us to
introduce relation $\mathbf{H} = curl \mathbf{A}$ strictly into the variational principle
(it follows after variation with respect to $\mathbf{H}$).

Supposing first that all variables introduced (including velocity) are independent we
obtain the set of variational equations of the form
\begin{equation}\label{mass_Er}
\delta \varphi \Longrightarrow \quad \partial_{t} \rho + div(\rho{\mathbf{v}}) = 0  ,
\end{equation}
\begin{equation}\label{VOL1B_Er}
\delta \rho  \Longrightarrow \quad  D\varphi =  w  - v^2/2  \, ,
    \end{equation}
\begin{equation}\label{mu_1_Er}
\delta  \boldit{\lambda} \Longrightarrow \quad   D \boldit{\mu} = 0 \, ,
\end{equation}
  \begin{equation}\label{lambda_Er}
\delta  \mu_{m}   \Longrightarrow \quad       \partial_{t} \lambda_{m} + div(\lambda_{m}
\mathbf{v}) =0  ,
   \end{equation}
\begin{equation}\label{s_eq_Er}
\delta  \sigma \Longrightarrow \quad  D s = 0 ,
\end{equation}
\begin{equation}\label{entr_Er}
\delta  s \Longrightarrow \quad     \partial_{t} \sigma + div(\sigma\mathbf{v}) =   -
\rho T ,
   \end{equation}
\begin{equation}\label{Magn_Er}
\delta  \mathbf{M}  \Longrightarrow \quad   \partial_{t}  \mathbf{A} =   [\mathbf{v} ,
curl \mathbf{A}] - \nabla \Lambda  ,
\end{equation}
  \begin{equation}\label{VOL1A_Er}
\delta   \mathbf{A}  \Longrightarrow \quad   \partial_{t}  \mathbf{M}  = \frac{curl
\mathbf{H}}{4\pi} + curl  [\mathbf{v}, \mathbf{M}]  .
      \end{equation}
\begin{equation}\label{DIV_Er}
\delta   \mathbf{H} \Longrightarrow \quad   \mathbf{H} = curl \mathbf{A} ,
\end{equation}
\begin{equation}\label{Er_1}
\delta  \Lambda \Longrightarrow \quad div \mathbf{M} = 0 ,
\end{equation}
where $w$ and $T$ are the enthalpy density and temperature.

Note that in this section we suppose the velocity field to be independent on other
variables. Therefore, variation with respect to $\mathbf{v}$ gives us the velocity
representation:
\begin{equation}\label{velocity_Er}
\delta   \mathbf{v}   \Longrightarrow \quad  \rho {\bf v} = - \rho \nabla \varphi -
{\lambda}_{m} \nabla {\mu}_{m} - \sigma \nabla s - \left[ {\bf H}, \mathbf{M} \right] .
\end{equation}
It is convenient to rewrite it in a shortened form that emphasizes it's structure.
Bearing in mind that the velocity potential $\varphi$, vector Lagrange markers
$\boldit{\mu}$, entropy $s$  and the vector potential $\mathbf{A}$ can be treated as
generalized coordinates, one can see that $\rho$, $\boldit{\lambda}$, $\sigma$ and
subsidiary field $\mathbf{M}$ are conjugated momenta, respectively. Let
\begin{equation}\label{5_09_02}
\mathcal{ Q} = (Q, \mathbf{A}) , \quad  Q = (\varphi, \boldit{\mu}, s) , \quad \mathcal{
P} = \delta \mathcal{A}/\delta \partial_{t}\mathcal{ Q} , \quad \mathcal{ P} = (P ,
\mathbf{M}) .
\end{equation}
Then the velocity representation takes the transparent form
\begin{equation}\label{5_09_02_1}
\mathbf{v} = \mathbf{v}_{0}(\mathcal{P}, \nabla \mathcal{Q}) , \quad \mathbf{v}_{0} =
\mathbf{v}_{h} + \mathbf{v}_{M} \, , \quad  \mathbf{v}_{h} = - \frac{P}{\rho} \nabla Q ,
\quad \mathbf{v}_{M} = - \frac{1}{\rho}\left[ {\bf H}, \mathbf{M} \right]  .
\end{equation}
Here subindexes $h$ and $M$ correspond to the "hydrodynamic" and "magnetic" parts of the
velocity field. The hydrodynamic part, $\mathbf{v}_{h}$, corresponds to the generalized
Clebsch representation, compare papers \cite{KK_97, KATS_01, KATS_02}, and magnetic part,
$\mathbf{v}_{M}$, coincides with the conventional term if we replace the divergence-free
field $\mathbf{M}$ by $curl \mathbf{S}$. The latter was first introduced by Zakharov and
Kuznetsov, cf. Ref. \cite{ZAK_KUZ_70}.

From the velocity representation Eq.~(\ref{5_09_02_1}) and the equations of motion
(\ref{mass_Er}) -- (\ref{VOL1A_Er}) it strictly follows that the velocity field
$\mathbf{v} = \mathbf{v}_{0}$ satisfies Euler equation with the magnetic force taken into
account. Namely, providing differentiation we obtain
\begin{equation}\label{Euler_Er}
  \rho D \mathbf{v}_{0} = - \nabla p + \frac{1}{4\pi} [curl \mathbf{H}, \mathbf{H}] ,
\end{equation}
where $p$ is the fluid pressure.

\subsection{Canonical variables.}\label{CAN_VAR}

 \hskip\parindent

The variational principle can be easily reformulated in the Hamiltonian form. Excluding
the magnetic and velocity fields by means of Eqs.~(\ref{DIV_Er}), (\ref{5_09_02_1}) we
arrive to the following Hamiltonian density
\begin{equation}\label{30_06_02D}
 \mathcal{H} = \mathcal{H}(\mathcal{P}, \nabla \mathcal{Q}) = \mathcal{P} \partial_{t}  \mathcal{Q} -  \mathcal{L}' =  \rho\frac{v_{0}^2}{2} +
\rho \varepsilon(\rho, s) + \frac{(rot \mathbf{A})^{2}}{8\pi} - (\mathbf{M}, \nabla
\Lambda)  .
\end{equation}
Equations of motion (\ref{mass_Er}) -- (\ref{VOL1A_Er}) can be expressed now in the
canonical form
\begin{equation}\label{30_06_07}
\partial_{t} {\cal Q} = {\delta {\cal H}_{V}}/{\delta { \cal P}} \, , \qquad
\partial_{t} {\cal P} = - {\delta {\cal H}_{V}}/ {\delta { \cal Q}}  \, ,
\qquad  {\cal Q} = (\varphi, \boldit{\mu}, s; \mathbf{A} ) \, , \quad  {\cal P} = (\rho,
\boldit{\lambda}, \sigma ; \mathbf{M} ) \, ;
\end{equation}
Eq.~(\ref{DIV_Er}) serves as a definition of magnetic field, and divergence-free
condition for the subsidiary field $\mathbf{M}$, Eq.~(\ref{Er_1}), follows from variation
of the action
\begin{equation}\label{Action_vol_Er_1}
\mathcal{A} =  \int d t  \int d \mathbf{r} \left( \mathcal{P}\partial_{t} \mathcal{Q} -
\mathcal{H}\right)
\end{equation}
with respect to $\Lambda$. Note that it is possible to put $\Lambda = 0$. Under this
assumption the divergence-free condition for the field $\mathbf{M}$ vanishes, but from
Eq.~(\ref{VOL1A_Er}) follows that $ div \mathbf{M}$ is conserved quantity, and supposing
that $div \mathbf{M} = 0$ holds for some initial moment we arrive to the conclusion that
this is valid for arbitrary moment. Nevertheless, it proves convenient to deal with
$\Lambda \ne 0$ that makes it possible to use different gauge conditions for the vector
potential.

The variational principle presented gives us the set of dynamic equations from which
follow conventional MHD equations, (\ref{mass_Er}), (\ref{entr_Er}), (\ref{Euler_Er}) and
equation for the frozen-in magnetic field, which follows from Eq.~(\ref{Magn_Er}) after
taking curl operation,
\begin{equation}\label{5_09_02_2}
\partial_{t}  \mathbf{H} =   curl [\mathbf{v} , \mathbf{H}] .
\end{equation}

On the contrary, if at some initial moment, $t = \bar{t}$, we have the conventional MHD
fields $\bar{\rho}$, $\bar{s}$, $\bar{\mathbf{v}}$ and  $\bar{\mathbf{H}}$, then we can
find the initial subsidiary fields $\bar{\varphi}$, $\bar{\boldit{\mu}}$,
$\bar{\boldit{\lambda}}$, $\bar{\sigma}$, $\bar{\mathbf{A}}$, $\bar{\mathbf{M}}$ and
$\bar{\Lambda}$, satisfying Eqs.~(\ref{DIV_Er}) -- (\ref{velocity_Er}). This can be done
up to the gauge transformations (do not changing the velocity and magnetic field) due to
the fact that the subsidiary fields play a role of generalized potentials. Then, if the
uniqueness conditions hold both for the conventional MHD equations and for the set of
variational equations, we arrive to the conclusion that corresponding solutions coincide
for all moments. In this sense we can say that these sets of equations are equivalent,
cf. Ref. \cite{KUZ_01}.

The complete representation of the velocity field in the form of the generalized Clebsch
representation, Eq.~(\ref{5_09_02_1}) allows, first, to deal with the MHD flows of
general type, including all types of breaks, cf. Ref. \cite{KATS_02}; second, for the
zero magnetic field it gives correct limit transition to the conventional hydrodynamics,
cf. Refs. \cite{KATS_01}, \cite{KK_97}; third, it allows to obtain the additional to the
known ones integrals and invariants of motion for the MHD flows: for instance,
generalized Ertel invariant, generalized vorticity and generalized helicity, see below.
The two last integrals were deduced for the particular case of incompressible flows in
the paper \cite{Vl95}.

Moreover, it is possible to show that representation (\ref{5_09_02_1}) is equivalent to
the one following from the Weber transformation, cf. Refs. \cite{Weber1868, lamb} and the
recent review \cite{Zh_Kuz_97}. The generalization of the Weber transformation for the
ideal MHD incompressible flows was obtained by Vladimirov and Moffatt, cf. Ref.
\cite{Vl95}.

\section{Generalized Weber transformation.}\label{WEB_TR}

\hskip\parindent

Suppose here that the fluid particles are labelled by Lagrange markers $\mathbf{a} =
(a_{1}, a_{2},a_{3})$. The label of the particle passing through point $\mathbf{r} =
(x_{1}, x_{2},x_{3})$ at time $t$ is then
\begin{equation}\label{W_1}
  \mathbf{a} = \mathbf{a}(\mathbf{r}, t)  ,
\end{equation}
and
\begin{equation}\label{W_2}
  D \mathbf{a} = \frac{\partial \mathbf{a}}{\partial t} + (\mathbf{v} \cdot \nabla )\mathbf{a} = 0 .
\end{equation}
The particle paths and velocities are given by the inverse function
\begin{equation}\label{W_3}
\mathbf{r} = \mathbf{r}(\mathbf{a} , t)   , \quad \mathbf{v} = D \mathbf{r} (\mathbf{a} ,
  t) =  \left. \frac{\partial \mathbf{r}}{\partial t}\right|_{\mathbf{a} = const}  .
\end{equation}
Let the initial position of the particle labelled $\mathbf{a}$ is $\mathbf{X}$, i.e.,
\begin{equation}\label{W_4}
 \mathbf{r}(\mathbf{a}, 0) = \mathbf{X}(\mathbf{a})   .
\end{equation}
A natural choice of label would be $\mathbf{X}(\mathbf{a}) = \mathbf{a}$; however it is
convenient to retain the extra freedom represented by the ``rearrangement function''
$\mathbf{X}(\mathbf{a})$.

We now seek to transform the equation of motion (\ref{Euler_Er}) to integrable form, by
generalization of the argument of Weber \cite{Weber1868}  (see, for example, Refs.
\cite{Serrin59}, \cite{Zh_Kuz_97}, and \cite{Vl95}. It is convenient to represent here
the equation of motion in the following form
\begin{equation}\label{Euler_W}
  D \mathbf{v} = - \nabla w + T \nabla s + [ \mathbf{J} , \mathbf{h} ]  ,
\end{equation}
where  $\mathbf{h} = \mathbf{H} /\rho$ and the vector $\mathbf{J}$ is defined according
to
\begin{equation}\label{Current}
  \mathbf{J} = \frac{curl \mathbf{H}}{4\pi}  ,
\end{equation}
being proportional to the current density. Multiplying Eq.~(\ref{Euler_W}) by $\partial
x_{k}/ \partial a_{i}$ we have
\begin{equation}\label{W_8}
(D v_{k}) \frac{\partial x_{k}}{\partial a_{i}} = - \frac{\partial w}{\partial x_{k}}
\frac{\partial x_{k}}{\partial a_{i}} + T \frac{\partial s}{\partial x_{k}}
\frac{\partial x_{k}}{\partial a_{i}} +  [ {\mathbf{J}}, {\mathbf{h}} ]_{k}
\frac{\partial x_{k}}{\partial a_{i}} \, .
\end{equation}
The l.h.s. can be represented in the form
\begin{equation}\label{W_9}
(D v_{k}) \frac{\partial x_{k}}{\partial a_{i}} = D \left(v_{k} \frac{\partial
x_{k}}{\partial a_{i}}\right) - \frac{\partial }{\partial a_{i}} (v^{2}/2) ,
\end{equation}
where  we have taken into account that  operator $D \equiv
\partial /\partial t |_{{\mathbf{a}}= const}$ and therefore $D x_{k} =
v_{k}$ and $D$ commute with derivative $\partial / \partial a_{i}$. Eq.~(\ref{W_8}) takes
now  the form
\begin{equation}\label{31_07_02_8}
D \left(v_{k} \frac{\partial x_{k}}{\partial a_{i}}\right)  = \frac{\partial }{\partial
a_{i}} (v^{2}/2 - w) + T \frac{\partial s}{\partial a_{i}} + [ {\mathbf{J}}, {\mathbf{h}}
]_{k} \frac{\partial x_{k}}{\partial a_{i}} \, .
\end{equation}
It is convenient to transform the last term  by means of the dynamical equation for the
subsidiary field $\mathbf{m} = \mathbf{M}/\rho$ (compare Eq.~(\ref{VOL1A_Er}))
\begin{equation}\label{31_07_02_9}
D \mathbf{m}   =  (\mathbf{m}, \nabla ) \mathbf{v} + \mathbf{J}/ \rho  .
\end{equation}
Then we can transform the last term in the r.h.s. of Eq.~(\ref{31_07_02_8}) to the form
of substantial derivative, see Appendix\label{???}

\begin{equation}\label{31_07_02_11}
[ {\mathbf{J}}, {\mathbf{h}} ]_{k} \frac{\partial x_{k}}{\partial a_{i}} =  D \left( [
{\mathbf{m}}, {\mathbf{H}} ]_{k}  \frac{\partial x_{k}}{\partial a_{i}} \right) .
\end{equation}

Analogously, the first two terms in the r.h.s. of Eq.~(\ref{31_07_02_8}) can be presented
as substantial derivatives by means of introducing subsidiary functions $\varphi$ and
$\sigma$, which satisfy equations (compare Eqs.~(\ref{entr_Er}), (\ref{VOL1B_Er}))
\begin{equation}\label{31_07_02_13}
D \left( \frac{\sigma}{\rho} \right)  = - T ,
\end{equation}
\begin{equation}\label{31_07_02_15}
D \varphi =  w - v^{2}/2 .
\end{equation}
Then
\begin{equation}\label{31_07_02_14}
T \frac{\partial s}{\partial a_{i}} = - \frac{\partial s}{\partial a_{i}} D \left(
\frac{\sigma}{\rho} \right) = - D \left( \frac{\partial s}{\partial
a_{i}}\frac{\sigma}{\rho} \right) , \quad \frac{\partial }{\partial a_{i}} (v^{2}/2 - w)
= - D \left( \frac{\partial  \varphi }{\partial a_{i}} \right) ,
\end{equation}
where we have taken into account that $D s = 0$ along with $D (\partial s / \partial
a_{i}) = 0$. Therefore, we can present the Euler equation (\ref{31_07_02_8}) in the
integrable form
\begin{equation}\label{31_07_02_17}
D \left(v_{k} \frac{\partial x_{k}}{\partial a_{i}}\right)  = - D \left( \frac{\partial
\varphi }{\partial a_{i}} \right) - D \left( \frac{\partial s}{\partial
a_{i}}\frac{\sigma}{\rho} \right) + D \left( [ {\mathbf{m}}, {\mathbf{H}} ]_{k}
\frac{\partial x_{k}}{\partial a_{i}} \right) .
\end{equation}
Integration leads to the relation
\begin{equation}\label{31_07_02_18}
v_{k} \frac{\partial x_{k}}{\partial a_{i}}  = -  \frac{\partial  \varphi }{\partial
a_{i}}  - \frac{\partial s}{\partial a_{i}}\frac{\sigma}{\rho}  - [\mathbf{H},
\mathbf{m}]_{k} \frac{\partial x_{k}}{\partial a_{i}} + b_{i} ,
\end{equation}
Here $\mathbf{b} = \mathbf{b}(\mathbf{a})$ does not depend on time explicitly, $D
\mathbf{b} = 0$, presenting the vector constant of integration. Multiplying this relation
by $\partial a_{i} /\partial x_{j}$ allows  reverting from Lagrangian variables
$(\mathbf{a}, t)$, to the Eulerian ones,  $(\mathbf{r}, t)$,
\begin{equation}\label{31_07_02_20}
\mathbf{v} = - \nabla \varphi + b_{k} \nabla  a_{k} -  \frac{\sigma}{\rho} \nabla s -
[\mathbf{h}, \mathbf{M}]   .
\end{equation}

This representation obviously coincides with the Clebsch representation obtained above
from the variational principle with constraints if one identifies $\mathbf{b}$ with
$-\boldit{\lambda}/\rho$ and $\mathbf{a}$ with $\boldit{\mu}$. Moreover, this proves
equivalence of description of the general type magnetohydrodynamic flows  in terms of
canonical variables introduced and the conventional description in Lagrange or Euler
variables. The equations of motion for the generalized coordinates and momenta follow now
from definitions of the subsidiary variables $\mathbf{a}$, $\mathbf{m} =
\mathbf{M}/\rho$, $\sigma$, $\varphi$ and $\mathbf{b}$.

Emphasize here  that the vector field $\mathbf{M} = \rho \mathbf{m}$ introduced by
Eq.~(\ref{31_07_02_9}) satisfies integral  relation
\begin{equation}\label{12_08_02_D_Er}
\partial_{t}  \int_{\Sigma} (\mathbf{M}, d \boldit{\Sigma}) = \int_{\Sigma} (\mathbf{J}, d
\boldit{\Sigma}) ,
\end{equation}
where $\boldit{\Sigma}$ presents some oriented area moving with the fluid. The proof of
this statement see in Appendix. Expressing $\mathbf{M} = curl \mathbf{S}$ and making use
of the Stocks theorem we arrive to a conclusion that time derivative of the vector
$\mathbf{S}$ circulation over the closed frozen-in contour $\partial \Sigma$ is
proportional to the current (remind, $\mathbf{J} = (4\pi)^{-1} curl \mathbf{H}$ and
differs from the current density by constant multiplier) intersecting the surface defined
by this contour,
\begin{equation}\label{12_08_02_D5_Er}
\partial_{t}  \int_{\partial \Sigma} (\mathbf{S}, d \mathbf{l}) = \int_{\Sigma} (\mathbf{J}, d
\boldit{\Sigma}) = (4\pi)^{-1} \int_{\partial \Sigma} (\mathbf{H}, d \mathbf{l})
\end{equation}
that highlights the physical meaning of the subsidiary field $\mathbf{S}$ usually
introduced for the canonical description of MHD flows. This fact was first indicated  in
Ref.~\cite{Vl95} for the incompressible flows. Now we see that it holds true for the
general case.

The vector constant of integration, $\mathbf{b}$,  may be expressed in terms of the
initial conditions,
\begin{equation}\label{31_07_02_21}
\begin{split}
b_{i} = \overline{V}_{k} (\mathbf{a}) \frac{\partial X_{k}}{\partial a_{i}}  +
\frac{\partial \varphi_{0} }{\partial a_{i}}
+ c_{0} \frac{\partial s}{\partial a_{i}} \, , \quad \quad \quad \quad \quad  \quad \quad  \quad \quad \\
\varphi_{0} = \varphi(\mathbf{a}, 0) , \quad c_{0} = \left( \frac{\sigma}{\rho} \right)
\biggl|_{t = 0}  ,  \quad \overline{V}_{k} (\mathbf{a}) = V_{k}
(\mathbf{a}) + [\mathbf{h}_{0}, \mathbf{M}_{0}]_{k} \, , \quad {V}_{k} (\mathbf{a}) = v_{k} (\mathbf{a}, 0) \, , \quad\quad \\
\mathbf{h}_{0} \equiv \mathbf{h}_{0}(\mathbf{a}) = \mathbf{h} (\mathbf{x}(\mathbf{a}, 0),
0) = \mathbf{h} (\mathbf{X}(\mathbf{a}) , 0) ,  \quad \mathbf{M}_{0} \equiv
\mathbf{M}_{0}(\mathbf{a}) = \mathbf{M} (\mathbf{x}(\mathbf{a}, 0), 0)  = \mathbf{M}
(\mathbf{X}(\mathbf{a}), 0) .
\end{split}
\end{equation}

Under special conditions, namely, for
\begin{equation}\label{1_08_02_1}
\mathbf{X}(\mathbf{a}) =  \mathbf{a} , \quad \mathbf{r}(\mathbf{a}, 0) = \mathbf{a}  ,
\quad \mathbf{a}(\mathbf{r}, 0) = \mathbf{r} ,
\end{equation}
from Eq.~(\ref{31_07_02_21}) follows
\begin{equation}\label{1_08_02_2}
b_{i} = \overline{V}_{i} (\mathbf{a})  + \frac{\partial  \varphi_{0} }{\partial a_{i}} +
c_{0} \frac{\partial s}{\partial a_{i}} \, .
\end{equation}
Adopting zero initial conditions,
\begin{equation}\label{1_08_02_3}
\mathbf{M}_{0}= 0 , \quad \varphi_{0} = 0  , \quad \sigma_{0} = 0  ,
\end{equation}
we obtain
\begin{equation}\label{2_08_02}
\mathbf{b} = \overline{\mathbf{V}}(\mathbf{a}) =  \widetilde{\mathbf{v}}(\mathbf{a}, 0)
\equiv  \widetilde{\mathbf{v}}_{0}(\mathbf{a}) = \mathbf{v}(\mathbf{a}, 0) \, ,
\end{equation}
where tilde indicates that we are dealing with the velocity field in the Lagrange
description, i.e., $\widetilde{\mathbf{v}}(\mathbf{a}, t)$ denotes velocity of the fluid
particle with label $\mathbf{a}$ at time $t$. Evidently,
$\widetilde{\mathbf{v}}(\mathbf{a}, t) = \mathbf{v}(\mathbf{r}, t)$, where $\mathbf{a}$
and $\mathbf{r}$ are linked by relations (\ref{W_1}) and (\ref{W_3})  for the specific
choice given by Eqs.~(\ref{1_08_02_1}), (\ref{1_08_02_3}). Then the velocity
representation takes the particular form
\begin{equation}\label{2_08_02_1}
\mathbf{v} = \mathbf{v}_{h} -  [\mathbf{h}, \mathbf{M}] , \quad \mathbf{v}_{h} \equiv -
\nabla \varphi + \widetilde{v}_{0k} \nabla  a_{k} -  \frac{\sigma}{\rho} \nabla s , \quad
\widetilde{\mathbf{v}}_{0}(\mathbf{r}) = \mathbf{v}(\mathbf{r}, 0) , \quad
\mathbf{a}(\mathbf{r} , 0) = \mathbf{r} .
\end{equation}
It differs from the one presented in Ref.~\cite{KUZ_01} by involving the entropy term.
Emphasize here that existence of this term allows to describe general type MHD and
hydrodynamic flows with arbitrary possible discontinuities, including shocks, slides and
rotational breaks, cf. Ref.~\cite{KK_97, KATS_01, KATS_02}. One can omit this term for
continuous barotropic and isentropic flows.

\section{Integrals of motion.}\label{INT_MOTION}

\hskip\parindent

The conservation laws, as it is well-known, follow from the specific symmetries of the
action. Existence of the relabelling transformations group (first discussed by Salmon in
Ref. \cite{Salmon82}) of the Lagrange markers, $\boldit{\mu}$, leads to the additional to
the energy, the fluid momentum and mass integrals of motion. These additional integrals
are expressed in terms of the Lagrange description of the motion, i.e., in terms of the
Lagrange markers, etc. Therefore, as a rule, they are  gauge dependent. The frozen-in
character of the magnetic field leads to the specific topological integrals of motion,
namely, magnetic helicity and, cross-helicity, first discussed by Moffatt in Ref.
\cite{Mof_69}, see also review \cite{Zh_Kuz_97}. Corresponding densities are respectively
\begin{equation}\label{8_09_02}
h_{M} = (\mathbf{A, H}) ,
\end{equation}
and
\begin{equation}\label{8_09_02_1}
h_{C} = (\mathbf{v, H}) .
\end{equation}
As it strictly follows from the dynamic equations, the local conservation law for the
magnetic helicity holds true for general type MHD flows
\begin{equation}\label{8_09_02_2}
\partial_{t} h_{M} + div \mathbf{q}_{M} = 0  , \quad \mathbf{q}_{M} =
\mathbf{v} h_{M} - \mathbf{H} \cdot \left( (\mathbf{A} , \mathbf{v} ) + \Lambda  \right)
.
\end{equation}
On the contrary, the cross-helicity in general case is governed by equation
    $$
\partial_{t} h_{C} /  \partial t = - div \left[\mathbf{v}  h_{C} +  ( w - v^{2}/2 ) \mathbf{H}
\right] + T div (s \mathbf{H})
    $$
and is not conserved. But for barotropic and isentropic flows the pressure $p = p(\rho)$
and $h_{C}$ is conserved:
\begin{equation}\label{8_09_02_3}
\partial_{t} h_{C} + div \mathbf{q}_{C} = 0 ,  \quad \mathbf{q}_{C} =
\mathbf{v}  h_{C} +  ( \chi - v^{2}/2 ) \mathbf{H} ,
\end{equation}
where $\chi = \int d p/\rho$.

For the general case there is known one more conserved quantity first discovered by
Gordin and Petviashvili, cf. Ref. \cite{Gordin87}. Corresponding density is
\begin{equation}\label{9_09_02}
h_{P} = (\mathbf{H}, \nabla s) ,
\end{equation}
and
\begin{equation}\label{9_09_02_1}
\partial_{t} h_{P} + div \mathbf{q}_{P} = 0 ,  \quad \mathbf{q}_{P} = \mathbf{v} h_{P} \, .
\end{equation}

With this local conserved quantity there is linked integral conservation law. Namely,
integrating $h_{P}$ over arbitrary substantial volume $\widetilde{V}$ we obtain conserved
quantity $\mathcal{I}_{P}$,
\begin{equation}\label{9_09_02_2}
\mathcal{I}_{P} = \int_{\widetilde{V}} d \mathbf{r} h_{P} \, , \quad \partial_{t}
\mathcal{I}_{P} = 0 .
\end{equation}

Note here that the latter quantity gives us example of the so called local Lagrange
invariants, cf. Refs. \cite{MCTYan_82, STY_90, TYan_93, VTY_95} and \cite{GP93,
Zh_Kuz_97}. By definition they obey the following equations
\begin{equation}\label{8_07_02_Er}
\partial_{t} \alpha + (\mathbf{v}, \nabla ) \alpha = 0 \, , \quad \partial_{t} {\mathbf{I}} + (\mathbf{v},
\nabla ) \mathbf{I} = 0 \, ,
\end{equation}
\begin{equation}\label{8_07_02_2_Er}
\partial_{t} {\mathbf{J}} + (\mathbf{v}, \nabla ) \mathbf{J} - (\mathbf{J}, \nabla ) \mathbf{v}  =
0 \, ,
\end{equation}
\begin{equation}\label{8_07_02_1_Er}
\partial_{t} {\mathbf{L}} + (\mathbf{v}, \nabla ) \mathbf{L} + (\mathbf{L}, \nabla ) \mathbf{v} +
[\mathbf{L}, curl \mathbf{v}] = 0  , \; {\mbox{or, equivalently,}}  \;
\partial_{t} {\mathbf{L}} + \nabla (\mathbf{v},  \mathbf{L}) - [\mathbf{v},
curl \mathbf{L}] = 0  .
\end{equation}
Here $\alpha$ and $\mathbf{I}$ present the scalar and vector  Lagrange invariants,
$\mathbf{J}$ is frozen-in field, and $\mathbf{L}$ presents $S$-type invariant in
terminology of Ref. \cite{TYan_93}, related to frozen-in surface. To these invariants it
is necessary to add the density $\rho$. Evidently, the quantity $h_{P}/\rho$ presents
$\alpha$-type invariant. The Lagrange markers $\boldit{\mu}$ and quantities
$\boldit{\lambda}/\rho$ give us examples of the vector Lagrange invariants, magnetic
field $\mathbf{H}$ divided by $\rho$, $\mathbf{h} = \mathbf{H}/\rho$ is invariant of the
$\mathbf{J}$- type, gradient of any scalar Lagrange invariant is $S$-type invariant,
\begin{equation}\label{9_09_02_5}
\mathbf{L}' = \nabla \alpha .
\end{equation}

There exist also another relations between different type invariants, see Refs.
\cite{GP93, Zh_Kuz_97}, allowing to produce new invariants. For instance, scalar product
of the $\mathbf{J}$ and $\mathbf{L}$ invariants presents some scalar Lagrange invariant,
symbolically
\begin{equation}\label{9_09_02_6}
\alpha' = (\mathbf{J}, \mathbf{L}) .
\end{equation}
The presented above invariant $h_{P}/\rho$ can be obtained by means of this relation if
we put $\mathbf{J} = \mathbf{h}$ and $\mathbf{L} = \nabla s$. Another examples present
relations generating $\mathbf{J}$- ($\mathbf{L}$)- type invariants by means of two
$\mathbf{L}$- ($\mathbf{J}$-) type invariants,
\begin{equation}\label{9_09_02_6A}
\mathbf{J}' = [\mathbf{L}, \mathbf{L}']/\rho ,
\end{equation}
\begin{equation}\label{28_09_02_1}
\mathbf{L}' = \rho [\mathbf{J}, \mathbf{J}'] .
\end{equation}

Note here that integrating of the density $h_{M}$ over arbitrary substantial volume does
not lead to the conserved integral. It is easy to check that
\begin{equation}\label{9_09_02_3}
\mathcal{I}_{M} = \int_{\widetilde{V}} d \mathbf{r} h_{M}
\end{equation}
satisfies relation
\begin{equation}\label{9_09_02_4}
\partial_{t} \mathcal{I}_{M} = \int_{\partial \widetilde{V}} d \Sigma
 \left( (\mathbf{A} , \mathbf{v} ) + \Lambda \right) H_{n}    \, ,  \quad H_{n} =
 (\mathbf{H,n}) \, ,
\end{equation}
where integration in the r.h.s. is performed over  the boundary $\partial \widetilde{V}$
of the volume $\widetilde{V}$, $\mathbf{n}$ is outward normal and $d \Sigma$ presents
infinitesimal area of the surface $\partial \widetilde{V}$. It is obvious that
$\mathcal{I}_{M}$ will be integral of motion if $H_{n}$ equals zero. This fact leads to
the conclusion that $\mathcal{I}_{M}$ presents integral of motion if we choose the
substantial volume in such a way that initial volume $\widetilde{V}|_{t = t_{0}}$ is such
that $H_{n}|_{t = t_{0}} =0$ because this condition is invariant of the motion: if
equality $H_{n} = 0$ holds for the initial moment then it holds true in the future.

Another way to make $\mathcal{I}_{M}$ invariant consists in fixing the gauge of the
vector potential $\mathbf{A}$ in such a way that $(\mathbf{A} , \mathbf{v} ) + \Lambda =
0$. Then the dynamic equation for $\mathbf{A}$, (\ref{Magn_Er}), takes  the form
    $$
\partial_{t} \mathbf{A}
+ \nabla (\mathbf{v},  \mathbf{A}) - [\mathbf{v}, curl \mathbf{A}] = 0 ,
    $$
i.e., $\mathbf{A}$ presents invariant of the $\mathbf{L}$- type. Under this gauge
condition the quantity $h_{M}/\rho$ presents the scalar Lagrange invariant, $D
(h_{M}/\rho) = 0$.

As for the local conservation law for the cross-helicity, Eq.~(\ref{8_09_02_3}), it
obviously leads to the integral conserved quantity $\mathcal{I}_{C}$ for the barotropic
flows but with following restriction: integration have to be performed over the specific
substantial volume such one that condition $H_{n}|_{\partial \widetilde{V}} = 0$ (this
condition is invariant of the motion) holds,
    $$
\partial_{t} \mathcal{I}_{C} = 0 \,  , \quad \mathcal{I}_{C} \equiv  \int_{\widetilde{V}} d \mathbf{r}
h_{C} \, , \quad H_{n}|_{\partial \widetilde{V}} = 0 .
    $$

Existence of the recursive procedure allowing one to construct new invariants on the
basis of the starting set of invariants, see Refs.~\cite{GP93, Zh_Kuz_97}, accentuates
the role of the local invariants among other conserved quantities. Although in terms of
the Lagrangian variables (such as the markers $\boldit{\mu}$) there exist a wide set of
invariants, see, for instance, Ref.~\cite{Zh_Kuz_97}, the most interesting invariants are
such that can be expressed in Eulerian (physical) variables and are gauge invariant.
Emphasize here that in the conventional hydrodynamics there exists Ertel invariant
$\alpha_{E}$,
\begin{equation}\label{11_09_02}
 \alpha_{E} = h_{E}/\rho , \quad h_{E} = (\boldit{\omega}, \nabla s) ,
\end{equation}
where $\boldit{\omega} = curl \mathbf{v}$ is vorticity,
\begin{equation}\label{11_09_02_1}
\partial_{t} h_{E} + div \mathbf{q}_{E} = 0 ,  \quad \mathbf{q}_{E} =  h_{E}\mathbf{v} , \quad D \alpha_{E} = 0
.
\end{equation}
Corresponding integral of motion reads
\begin{equation}\label{12_09_02}
\partial_{t} \mathcal{I}_{E} =  0 \,  ,
\quad \mathcal{I}_{E} \equiv  \int_{\widetilde{V}} d \mathbf{r} h_{E} \, .
\end{equation}
Note here that $\mathcal{I}_{E} = 0$ holds true for arbitrary substantial volume $
\widetilde{V}$.

The Ertel invariant density has the structure of the Eq.~(\ref{9_09_02_6}) with
$\mathbf{L} = \nabla s$, $\mathbf{J} = \boldit{\omega}/\rho$, where $\boldit{\omega}$ is
vorticity, $\boldit{\omega} = curl \mathbf{v}$ (remind that $\boldit{\omega}$ is a
frozen-in field for the barotropic hydrodynamic flows). In the hydrodynamic case there
exists also the helicity invariant
\begin{equation}\label{12_09_02_1}
h_{H} = (\boldit{\omega}, \mathbf{v}) ,
\end{equation}
which has topological meaning, defining  knottness of the flow. It satisfies equation
\begin{equation}\label{12_09_02_2}
\partial_{t} h_{H} + div \mathbf{q}_{H} = 0 , \quad \mathbf{q}_{H} = h_{H} \mathbf{v} + (\chi -
v^{2}/2) \boldit{\omega}  ,
\end{equation}
and evidently leads to the corresponding integral conservation law
\begin{equation}\label{12_09_02_3}
\partial_{t} \mathcal{I}_{H} =  0 \,  ,
\quad  {\mbox{for}} \quad  \omega_{n}|_{\partial \widetilde{V}} = 0 \, , \quad
\mathcal{I}_{H} \equiv  \int_{\widetilde{V}} d \mathbf{r} h_{E} \,  .
\end{equation}

For the MHD case the vector $\boldit{\omega}/\rho$ does not present frozen-in field due
to the fact that magnetic force is not potential. It seems rather evident that for the
MHD case there have to exist integrals of motion generalizing the conventional helicity
and Ertel invariant along with vorticity one, which have to pass into conventional ones
for vanishing magnetic field. The generalization for the vorticity and helicity
invariants was obtained by authors of the paper \cite{Vl95} for the particular case of
the incompressible flows. In the following section it is shown that there exists MHD
generalization for the Ertel invariant, and results of the paper \cite{Vl95} relating to
the vorticity and helicity can be extended for incompressible barotropic MHD flows.

\subsection{Generalized vorticity.}\label{Gen_vorticity}

\hskip\parindent

Let us prove that the quantity $\boldit{\omega}_{h}/\rho$, where
\begin{equation}\label{12_09_02_4}
\boldit{\omega}_{h} \equiv curl \mathbf{v}_{h}  = - \left[\nabla
\left(\frac{P}{\rho}\right), \nabla Q\right] = -  \left[\nabla
\left(\frac{\lambda_{m}}{\rho}\right), \nabla \mu_{m}\right] - \left[\nabla
\left(\frac{\sigma}{\rho}\right), \nabla s\right] ,
\end{equation}
presents frozen-in field  (`hydrodynamic' part of the vorticity) for the barotropic MHD
flows. It would be trivial consequence of the fact that $[\mathbf{L}, \mathbf{L}']/\rho$,
where $\mathbf{L}$, $\mathbf{L}'$ are Lamb type invariants, is local invariant of the
frozen-in type if all quantities $Q$ and $P/\rho$ satisfy homogeneous transport equations
being $\alpha$- or $\mathbf{I}$ type invariants (remember, that $\nabla \alpha$ and
$\nabla I_{m}$ are $\mathbf{L}$ type invariants). But $\varphi$ and $\sigma/\rho$ satisfy
inhomogeneous equations of motion. Therefore, let us start with equation of motion for
the `hydrodynamic' part of the velocity. Differentiating representation (\ref{5_09_02_1})
and making use of relations
    $$
D (\nabla X) = \nabla (D  X) - (\nabla v_{m}) \cdot \partial_{m} X
    $$
we have
    $$
D \mathbf{v}_{h} = - D \left(\frac{P}{\rho}\right) \cdot \nabla Q - \frac{P}{\rho} \cdot
\nabla (D Q) + \frac{P}{\rho} (\nabla v_{m}) \cdot \partial_{m} Q = T \nabla s - \nabla
(w - v^{2}/2) - v_{hm}(\nabla v_{m}) \, ,
    $$
or, after simple rearrangements,
\begin{equation}\label{12_09_02_5}
D \mathbf{v}_{h} = - \nabla p /\rho + (v_{m} - v_{hm}) \cdot \nabla v_{m}  \, .
\end{equation}
Taking the curl of this equation leads to
    $$
\partial_{t} \boldit{\omega}_{h} = - curl ((v_{m} \partial_{m} ) \mathbf{v}_{h}) + [\nabla \rho, \nabla
p]/\rho^{2} - curl (v_{hm} \nabla v_{m}) =
    $$
    $$
= [\nabla \rho, \nabla p]/\rho^{2} + curl [(v_{m} \nabla v_{hm}) - (v_{hm} \nabla v_{m})]
\, .
    $$
The term in the square brackets is equal to $[(v_{m} \nabla v_{hm}) - (v_{hm} \nabla
v_{m})] = [\mathbf{v}, \boldit{\omega}_{h}]$ and we obtain
\begin{equation}\label{12_09_02_6}
\partial_{t} \boldit{\omega}_{h} = [\nabla \rho, \nabla p]/\rho^{2}  + curl [\mathbf{v},
\boldit{\omega}_{h}]\, .
\end{equation}
For barotropic flows the first term in the r.h.s. becomes zero and we can see that
$\boldit{\omega}_{h}/\rho$ is frozen-in field,
\begin{equation}\label{12_09_02_6A}
D \left(\frac{\boldit{\omega}_{h}}{\rho} \right) = \left(\frac{\boldit{\omega}_{h}}{\rho}
, \nabla \right)\mathbf{v} \, .
\end{equation}
For $\mathbf{H} = 0$ $\boldit{\omega}_{h}$ corresponds to the conventional hydrodynamic
vorticity.

In spite of the gauge dependence of the generalized vorticity, it frozenness gives us
possibility to introduce the generalized helicity integral of motion.

\subsection{Generalized helicity.}\label{Gen_helicity}

\hskip\parindent

Now we can prove that generalized helicity, $h_{H}$, defined in terms of the
`hydrodynamic' part of the velocity,
\begin{equation}\label{13_09_02}
h_{H} = (\boldit{\omega}_{h} , \mathbf{v}_{h}) ,
\end{equation}
is integral of motion for barotropic flows. Namely, differentiating Eq.~(\ref{13_09_02})
and taking for account Eqs.~(\ref{12_09_02_5}), (\ref{12_09_02_6}) we arrive for the
barotropic flows to the local conservation law of the form (rather cumbersome
calculations are presented in Appendix):
\begin{equation}\label{13_09_02_1}
\partial_{t} h_{H} + div \mathbf{q}_{H} = 0 , \quad \mathbf{q}_{H} =
h_{H}\mathbf{v} + (\chi - v^{2}/2)\boldit{\omega}_{h}   \, .
\end{equation}
In analogy with the hydrodynamic case we arrive to the conclusion that the integral
helicity $\mathcal{I}_{H}$ (defined  by means of Eq.~(\ref{12_09_02_3})) is integral
invariant, moving together with the fluid if the normal component of the vorticity tends
zero, $\omega_{hn} = 0$, on the surface of the corresponding substantial volume
$\widetilde{V}$. Note that the condition $\omega_{hn} = 0$ is invariant of the flow (due
to the frozen--in character of $\boldit{\omega}_{h}/\rho$) and therefore it can be
related to the initial surface only.

\subsection{Generalized Ertel invariant.}

\hskip\parindent

Let us show here that there exists strict generalization of the Ertel invariant for the
MHD case. For this purpose let us prove that without any restrictions related to the
character of the flow the quantity
\begin{equation}\label{13_09_02_2}
h_{E} = (\boldit{\omega}_{h}, \nabla s)
\end{equation}
satisfies conservation law of the form
\begin{equation}\label{15_09_02}
\partial_{t} h_{E} + div \mathbf{q}_{E} = 0 , \quad \mathbf{q}_{E} = h_{E} \mathbf{v} .
\end{equation}
Equivalently, the quantity $\alpha_{E} = h_{E}/\rho$ is transported by the fluid
\begin{equation}\label{15_09_02_1}
D  \alpha_{E} = 0 , \quad \alpha_{E} = h_{E}/\rho ,
\end{equation}
presenting $\alpha$- type invariant. For the barotropic flows it immediately follows from
the fact that $\boldit{\omega}_{h}/\rho$ presents frozen-in field if one takes for
account  the composition rules given by Eqs.~(\ref{9_09_02_6}) and (\ref{9_09_02_5}). In
order to make the proof for the non barotropic flows more transparent let us consider
something more general situation. Let $\widetilde{\mathbf{J}}$ satisfy equation of motion
of the form
\begin{equation}\label{15_09_02_2}
D  \widetilde{\mathbf{J}}  =  (\widetilde{\mathbf{J}}, \nabla) \mathbf{v} + \mathbf{Z} ,
\end{equation}
differing from the frozen field equation (\ref{8_07_02_2_Er}) by existence of the term
$\mathbf{Z}$ that violates homogeneity. Then, if $\alpha$ represents any scalar Lagrange
invariant, we have
    $$
D  (\widetilde{\mathbf{J}} , \nabla \alpha) = \left( D \widetilde{\mathbf{J}} , \nabla
\alpha \right) + \left( \widetilde{\mathbf{J}}, D(\nabla \alpha) \right)
 = \left(    \mathbf{Z} , \nabla \alpha
\right) + \left( ((\widetilde{\mathbf{J}} , \nabla)  \mathbf{v} ), \nabla \alpha \right)
- \left( \widetilde{\mathbf{J}}, (\nabla v_{m}) \cdot \partial_{m} \alpha ) \right) .
$$
Here the two last terms cancel and we get
\begin{equation}\label{15_09_02_3}
D  (\widetilde{\mathbf{J}} , \nabla \alpha )  = \left(    \mathbf{Z} , \nabla \alpha
\right) \quad {\mbox{if}}  \quad  D  \widetilde{\mathbf{J}}  =  (\widetilde{\mathbf{J}},
\nabla) \mathbf{v} + \mathbf{Z} , \quad {\mbox{and}}  \quad  D \alpha = 0 .
\end{equation}
For $\mathbf{Z} = 0$ this relations prove the generating rule of Eq.~(\ref{9_09_02_6}).
But we can see that $(\widetilde{\mathbf{J}} , \nabla \alpha )$ will present the local
Lagrange invariant under more restrictive condition $( \mathbf{Z} , \nabla \alpha) = 0$.
That is the case for the Ertel invariant: $\mathbf{Z} = [\nabla \rho, \nabla p]/\rho^{3}$
is orthogonal to $\nabla s$ due to the fact that the scalar product of any three
thermodynamic quantities is equal zero (because any thermodynamic variable in the
equilibrium state can be presented as function of two basic variables). This ends the
proof.

The conserved integral quantity associated with $\alpha_{E}$ is
\begin{equation}\label{15_09_02_4}
\mathcal{I}_{E} = \int_{ \widetilde{V}} d \mathbf{r} h_{E} \, ,  \quad
\partial_{t} \mathcal{I}_{E} = 0  .
\end{equation}
Note here that by the structure $\mathcal{I}_{E}$ is not gauge invariant in contrast to
the hydrodynamic case. Let us examine it change under gauge transformation changing
$\mathbf{v}_{h} \Rightarrow \mathbf{v}'_{h}$, $\mathbf{v}_{M} \Rightarrow
\mathbf{v}'_{M}$ with
    $$
\mathbf{v}'_{h} + \mathbf{v}'_{M} = \mathbf{v}_{h} + \mathbf{v}_{M} \, .
    $$
Then
    $$
\mathcal{I}'_{E} - \mathcal{I}_{E}  = \int_{ \widetilde{V}} d \mathbf{r} (\nabla s,
\boldit{\omega}'_{h} - \boldit{\omega}_{h}) =  \int_{ \widetilde{V}} d \mathbf{r} (\nabla
s, \boldit{\omega}_{M} - \boldit{\omega}'_{M}) \, .
    $$
But $(\nabla s, \boldit{\omega}_{M} - \boldit{\omega}'_{M}) = - div [\nabla s,
(\mathbf{v}'_{M} - \mathbf{v}_{M})]$ and therefore we can proceed as follows
    $$
\mathcal{I}'_{E} - \mathcal{I}_{E}  =   - \int_{ \partial \widetilde{V}} d \Sigma
(\mathbf{n}, [\nabla s, (\mathbf{v}'_{M} - \mathbf{v}_{M})] ) \, .
    $$
Taking into account that $\mathbf{v}'_{M} - \mathbf{v}_{M} = - [\mathbf{h}, \mathbf{M}' -
\mathbf{M} ]$ we obtain
\begin{equation}\label{15_09_02_5}
\mathcal{I}'_{E} - \mathcal{I}_{E}  = \int_{ \partial \widetilde{V}} d \Sigma
\left(\mathbf{n}, [\nabla s, [\mathbf{h}, \mathbf{M}' - \mathbf{M} ]] \right) .
\end{equation}
Inasmuch as both $\mathbf{M}'$ and $\mathbf{M}$ satisfy Eq.~(\ref{VOL1A_Er}), their
difference  is governed by homogeneous equation
    $$
 \partial_{t}  \overline{\mathbf{M}}  = curl  [\mathbf{v}, \overline{\mathbf{M}}]  ,
    $$
i.e. $\overline{\mathbf{m}} = \overline{\mathbf{M}}/\rho$ is frozen-in field. Then we
arrive to the conclusion that the vector $[\nabla s, [\mathbf{h}, \overline{\mathbf{m}}]]
$ entering the integrand presents frozen-in field, as it follows from recursive relations
Eqs.~(\ref{9_09_02_5}) -- (\ref{9_09_02_6A}). Therefore, if we adopt relation
$(\mathbf{n}, [\nabla s, [\mathbf{h}, \overline{\mathbf{m}}]])|_{\partial \widetilde{V}}
= 0$  as initial condition, then it holds true for all moments. But we cannot choose the
(initial) substantial volume in such a way that relation
\begin{equation}\label{16_09_02}
(\mathbf{n}, [\nabla s, [\mathbf{h}, \mathbf{m}']])|_{\partial \widetilde{V}} =
(\mathbf{n}, [\nabla s, [\mathbf{h}, \mathbf{m}]])|_{\partial \widetilde{V}}
\end{equation}
holds true for any change of the gauge. Thus integral Ertel invariant is gauge dependent.
Nevertheless, we can point out some subset of the gauge transformations under which
$\mathcal{I}_{E}$ is invariant. Namely, let $\mathbf{M}|_{t = t_{0}} = f \mathbf{H}$ for
some initial moment, $t = t_{0}$, where $f$ have to satisfy condition $(\mathbf{H},
\nabla f ) =0 $, following from the divergence--free character of $\mathbf{M}$. Then
relation (\ref{16_09_02}) fulfills for the initial moment and therefore it holds true  at
all moments also. The specific choice $f = 0$ leads to additional restriction for the
gauge transformations but it is convenient due to its simplicity. Summarizing, we can say
that the Ertel invariant is partly gauge independent.

\newpage 

\section{Conclusions.}\label{Conlusions}

\hskip\parindent

The results obtained can be summarized as follows. First, there is presented variant of
introducing the canonical description of the MHD flows by means of the variational
principle with constraints. It is shown that in order to describe general type MHD flows
it is necessary to use in the generalized Clebsch type representation of the fluid
velocity field vector Clebsch variables (the Lagrange markers and conjugate momenta)
along with the entropy term (compare papers \cite{KK_97, KATS_01} describing hydrodynamic
case) and the conventional magnetic term introduced first in the paper \cite{ZAK_KUZ_70}.
Such complete representation allows one to deal with general type MHD flows, including
all type of breaks, see Ref. \cite{KATS_02}. Second, it is proved that introduced in the
paper generalized Weber transformation leads to the velocity representation, which
equivalent to the one introduced by means of the variational principle. Third, there is
proved existence of the generalized Ertel invariant for MHD flows. Forth, there are
generalized the vorticity and helicity invariants for the compressible barotropic MHD
flows (first discussed for the incompressible case in cf. \cite{Vl95}). Fifth, the
relations between the local and integral invariants are discussed along with the gauge
dependence of the latter.

As a consequence of the completeness of the representation proposed we arrive to the
correct limit transition from the MHD to conventional hydrodynamic flows. The results
obtained allow one to deal with the complicated MHD problems by means of the Hamiltonian
variables. The use of such approach was demonstrated for the specific case of
incompressible flows in the series of papers \cite{Vl95, Vl96}  devoted to the nonlinear
stability criteria. Emphasize, that existence of the additional invariants proved in our
paper is of very importance for the stability problems.

Note here that existing of the additional basic invariants of the motion makes it actual
to examine the problem of the complete set of independent invariants, cf.
\cite{Zh_Kuz_97}. This problem needs special discussion together with related problem of
their gauge invariance. One more open problem is connected with the great number of the
generalized coordinates and momenta involved in the approach discussed. Here the question
arises if it is possible to reduce this number without loosing the generality.

\section*{Appendix A}

\hskip\parindent

In order to prove Eq.~(\ref{31_07_02_11}) let us substitute $\mathbf{J}$ from
Eq.~(\ref{31_07_02_9}) into expression $[ {\mathbf{J}}, {\mathbf{h}} ]_{k} {\partial
x_{k}}/{\partial a_{i}}$. Then
\begin{equation}\label{31_07_02_10EA}
\begin{split}
[ {\mathbf{J}}, {\mathbf{h}} ]_{k} \frac{\partial x_{k}}{\partial a_{i}} = [
D{\mathbf{m}}, {\mathbf{H}} ]_{k} \frac{\partial x_{k}}{\partial a_{i}} - [ (\mathbf{m},
\nabla ) \mathbf{v} , {\mathbf{H}} ]_{k} \frac{\partial
x_{k}}{\partial a_{i}} = \\
= \frac{\partial x_{k}}{\partial a_{i}} D \left( [ {\mathbf{m}}, {\mathbf{H}} ]_{k}
\right) -  \left( [ {\mathbf{m}}, D ( \rho \mathbf{h}) ]_{k} + [ (\mathbf{m}, \nabla )
\mathbf{v} , {\mathbf{H}} ]_{k} \right) \frac{\partial x_{k}}{\partial a_{i}} .
\end{split}
\end{equation}
Proceeding with the terms in the second brackets we obtain
  \begin{equation}\label{28_09_02}
\begin{split}
[ {\mathbf{m}}, D ( \rho \mathbf{h}) ]_{k} + [ (\mathbf{m}, \nabla ) \mathbf{v} ,
{\mathbf{H}} ]_{k} = [ {\mathbf{m}}, \mathbf{h} ]_{k} \cdot D  \rho + [ \rho \mathbf{m},
D   \mathbf{h} ]_{k}  + [ (\mathbf{m}, \nabla ) \mathbf{v} , {\mathbf{H}} ]_{k} = \\ =  -
[ \mathbf{M} , \mathbf{h} ]_{k} \cdot div \mathbf{v} + [ \mathbf{M}, (\mathbf{h} ,
\nabla) \mathbf{v} ]_{k}  + [ (\mathbf{M}, \nabla ) \mathbf{v} , {\mathbf{h}} ]_{k}  =  -
[\mathbf{M}, \mathbf{h}]_{s}
\partial_{k} v_{s} \, ,
\end{split}
\end{equation}
where $\mathbf{M} = \rho \mathbf{m}$ and there are is taken for account dynamic equation
$D \mathbf{h} = (\mathbf{h}, \nabla ) \mathbf{v}$ and identity
$$
[ \mathbf{M}, (\mathbf{h} , \nabla) \mathbf{v} ]_{k} + [ (\mathbf{M}, \nabla ) \mathbf{v}
, {\mathbf{h}} ]_{k} = [\mathbf{M}, \mathbf{h}]_{k} \partial_{s} v_{s} - [\mathbf{M},
\mathbf{h}]_{s} \partial_{k} v_{s}  \, .
$$
Introducing for brevity notation
    $$
\mathbf{Y} = [\mathbf{m}, \mathbf{H}] \equiv [\mathbf{M}, \mathbf{h}] ,
    $$
we can represent the r.h.s. of Eq.~(\ref{31_07_02_10EA}) as
    $$
\frac{\partial x_{k}}{\partial a_{i}} \cdot D Y_{k}  +  Y_{s} \frac{\partial
x_{k}}{\partial a_{i}}
\partial_{k} v_{s} = \frac{\partial x_{k}}{\partial a_{i}} \cdot D Y_{k}  + Y_{s} \frac{\partial v_{s}}{\partial
a_{i}} = \frac{\partial x_{k}}{\partial a_{i}} \cdot D Y_{k}  + Y_{s}
\frac{\partial}{\partial a_{i}} (D x_{s} ) = D \left(Y_{k} \frac{\partial x_{k}}{\partial
a_{i}} \right) \, .
    $$
This proves Eq.~(\ref{31_07_02_11}).

Let us check up now the integral relation (\ref{12_08_02_D_Er}). It is sufficient to
prove the differential form, namely
\begin{equation}\label{12_08_02_D_Er_PR}
D (\mathbf{M}, d \boldit{\Sigma}) = (\mathbf{J}, d \boldit{\Sigma}) ,
\end{equation}
where $d \boldit{\Sigma}$ presents some infinitesimal oriented area moving with the
fluid. It can be presented in the form
\begin{equation}\label{12_08_02_D1_ER}
 d \boldit{\Sigma}  = [d \mathbf{l}_{1}, d \mathbf{l}_{2}] ,
\end{equation}
where $d \mathbf{l}_{1}$, $d \mathbf{l}_{2}$ are frozen--in linear elements. Thus, $d
\mathbf{l}_{a}$, $a = 1,2$, are invariants of the $\mathbf{J}$ type and satisfy equations
    $$
D ( d \mathbf{l}_{a}) = (d \mathbf{l}_{a}, \nabla) \mathbf{v} .
    $$
Consequently, from the recursion relation Eq.~(\ref{28_09_02_1}) it follows that $\rho d
\boldit{\Sigma}$ is $\mathbf{L}$- type invariant and therefore is governed by dynamic
equation of the form:
    $$
D (\rho  d \boldit{\Sigma}) = - \nabla (\rho \mathbf{v} d \boldit{\Sigma}) + [\mathbf{v},
curl  (\rho  d \boldit{\Sigma}) ] ,
    $$
or in the coordinates,
\begin{equation}\label{28_09_02_2}
 D (\rho  d {\Sigma}_{i}) = - (\rho  d {\Sigma}_{k}) \partial_{i} v_{k}  \, .
\end{equation}
Now it is easy to prove relation (\ref{12_08_02_D_Er_PR}) without any restrictions for
the type of flow. Namely,
\begin{equation}\label{28_09_02_3}
\begin{split}
D (\mathbf{M}, d \boldit{\Sigma} ) = D (\mathbf{m}, \rho d \boldit{\Sigma} ) = (D
\mathbf{m}, \rho d \boldit{\Sigma} ) +   m_{i} D (\rho d \Sigma_{i} ) = \\ = (\rho d
\boldit{\Sigma}, (\mathbf{m}, \nabla ) \mathbf{v} ) + (\mathbf{J}, d \boldit{\Sigma}) -
m_{i} \rho d \Sigma_{k}
\partial_{i} v_{k} =  (\mathbf{J}, d \boldit{\Sigma}) \, .
\end{split}
\end{equation}

In order to prove the helicity conservation, Eq.~(\ref{13_09_02_1}), let us consider some
scalar quantity of the form
    $$
Y = (\mathbf{v}_{h} \mathbf{J}) ,
    $$
where $\mathbf{J}$ is frozen-in field. Then, taking for account that
Eq.~(\ref{12_09_02_5})  for the barotropic flows can be presented in the form
    $$
D \mathbf{v}_{h} = - \nabla ( \chi - v^{2}/2)  - v_{hm} \cdot \nabla v_{m} \, , \quad
\chi \equiv \int d p /\rho ,
    $$
we obtain
    $$
D Y = (D \mathbf{v}_{h}, \mathbf{J}) + (\mathbf{v}_{h}, D \mathbf{J}) = - \left( \nabla (
\chi - v^{2}/2), \mathbf{J} \right) .
    $$
For $\mathbf{J} = \boldit{\omega}_{h}/\rho $ we proceed
    $$
D (\mathbf{v}_{h}, \boldit{\omega}_{h}/\rho) =  - \rho^{-1} \left( \nabla ( \chi -
v^{2}/2), \boldit{\omega}_{h} \right) = - \rho^{-1} div \left(   (\chi - v^{2}/2)
\boldit{\omega}_{h}  \right) \, .
    $$
Then
    $$
D (\mathbf{v}_{h}, \boldit{\omega}_{h}) = \rho D (\mathbf{v}_{h},
\boldit{\omega}_{h}/\rho) + (\mathbf{v}_{h}, \boldit{\omega}_{h}/\rho) D \rho = - div
\left(   (\chi - v^{2}/2) \boldit{\omega}_{h}  \right)  - \left(\mathbf{v}_{h},
\boldit{\omega}_{h} \right) div \mathbf{v} ,
    $$
or
\begin{equation}\label{30_09_02_A1}
\partial_{t} (\mathbf{v}_{h}, \boldit{\omega}_{h}) = - div \mathbf{q}_{h} \, , \quad
\mathbf{q}_{h} = (\chi - v^{2}/2) \boldit{\omega}_{h} + \mathbf{v} \left(\mathbf{v}_{h},
\boldit{\omega}_{h} \right)
\end{equation}
that evidently coincides with Eq.~(\ref{13_09_02_1}).

It is noteworthy that the proof is valid for any $\mathbf{J}$- type invariant if the
field $\rho\mathbf{J}$  is divergence--free:
\begin{equation}\label{30_09_02_A2}
\partial_{t} (\rho \mathbf{J}, \mathbf{v}_{h} ) = - div \mathbf{q} \, , \quad
\mathbf{q} = (\chi - v^{2}/2) \rho \mathbf{J} + \mathbf{v} \left(\rho \mathbf{J},
\mathbf{v}_{h} \right) \quad {\mbox{for}} \quad div (\rho\mathbf{J}) = 0 .
\end{equation}
For instance,  choosing  $\mathbf{J} = \mathbf{h}$ immediately leads to cross--helicity
invariant if one takes for account that $(\mathbf{H}, \mathbf{v}_{h}) = (\mathbf{H},
\mathbf{v})$.

\subsection*{Acknowledgment}

\frenchspacing

This work was supported by the INTAS (Grant No. 00-00292).

\medskip

\end{document}